\documentclass[a4paper]{article}
\usepackage[a4paper,includemp=false,body={16cm,24.7cm},vcentering]{geometry}
\usepackage[latin1]{inputenc}
\usepackage[T1]{fontenc}
\usepackage{mathptmx} 
\usepackage[bf,center]{caption}
\usepackage{graphicx} 

\usepackage{url,amsmath,amssymb,amsthm}

\makeatletter
\renewcommand{\section}{\@startsection{section}{1}{0mm}{30pt}{12pt}{\normalfont\normalsize\bfseries}}

\renewcommand{\subsection}{\@startsection{subsection}{2}{0mm}{18pt}{12pt}{\normalfont\normalsize\itshape}}
\makeatother

\newcommand{\Title}[1]{\begin{center}{\bfseries\fontsize{12pt}{12pt}\selectfont#1}\end{center}}
\newcommand{\Author}[2]{\begin{center}{\fontsize{12pt}{12pt}\selectfont#1}\\{\it #2~}\end{center}}
\newcommand{\Introduction}{\section*{Introduction}}
\newcommand{\Conclusion}{\section*{Conclusion}}

\begin{document}

\Title{Interferometry, spectroscopy and astrometry of the bright eclipsing system $\delta$\,Velorum}

\Author{P. Kervella$^1$, A. M\'erand$^2$, T. Pribulla\,$^3$}{
1. LESIA, Observatoire de Paris, CNRS\,UMR\,8109, UPMC, Universit\'e Paris Diderot,
5 place Jules Janssen, 92195 Meudon, France, pierre.kervella@obspm.fr\\
2. European Southern Observatory, Alonso de Cordova 3107, Casilla 19001, Vitacura, Santiago 19, Chile\\
3. Astronomical Institute Slovak Academy of Sciences 059 60 Tatransk\'a Lomnica, Slovak Republic}

\Introduction

\noindent

The bright southern star $\delta$ Vel is a multiple system comprising at least three stars. Its brightest component, Delta Vel A, was identified in 2000 as one of the brightest eclipsing system in the sky. Its eclipses are easily observable with the unaided eye, a remarkable property shared only by Algol, $\beta$\,Aur, $\alpha$\,CrB and $\psi$\,Cen. We determined dynamical masses from a combination of spectroscopy, high-precision astrometry of the orbits of Aab-B and Aa-Ab using adaptive optics (VLT/NACO) and optical interferometry (VLTI/AMBER). The main eclipsing component is a pair of A-type stars in rapid rotation. We modeled the photometric and radial velocity measurements of the eclipsing pair Aa-Ab using a self consistent method based on physical parameters (mass, radius, luminosity, rotational velocity). From this modeling, we derive the fundamental parameters of the eclipsing stars with a typical accuracy of 1\%. We find that they have similar masses, respectively $2.43 \pm 0.02$ and $2.27 \pm 0.02\,M_\odot$. The physical parameters of the tertiary component ($\delta$\,Vel~B) are also derived, although to a lower accuracy, as well as the parallax of the system, $\pi = 39.8 \pm 0.4$\,mas. This value is in satisfactory agreement ($-1.2\,\sigma$) with the Hipparcos parallax of the system ($\pi_{\rm Hip} =40.5 \pm 0.4$\,mas).


\section{Overall properties of the system}

\begin{figure}[ht]
\centering
\includegraphics[height=5cm]{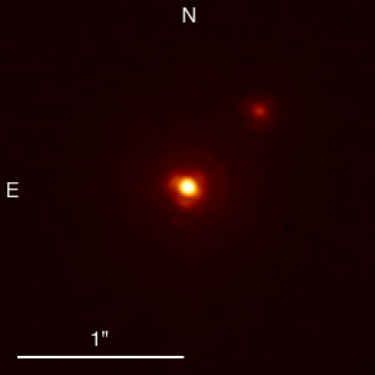} \hspace{8mm}
\includegraphics[height=5cm]{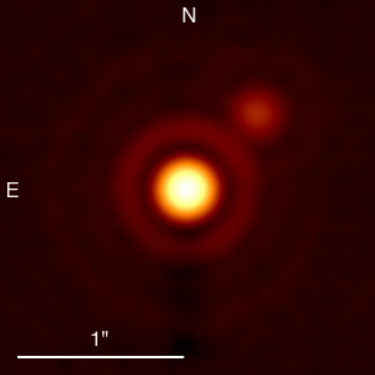}
\caption{VLT/NACO (2.17\,$\mu$m, left) and VLT/VISIR (8.6\,$\mu$m, right) images of $\delta$\,Vel~AB.  \label{nacovisir}}
\end{figure}

With $m_V=1.96$, $\delta$ Vel (HD 74956, HIP 41913, GJ 321.3, GJ 9278) is one of the brightest multiple stellar system in the southern sky. This object has several interesting observational properties. Firstly, it was reported in 2000 that $\delta$\,Vel hosts one of the brightest eclipsing binaries in the sky (\cite{otero00}), with a remarkably long orbital period ($P \approx 45$\,days). This eclipsing binary is also one of the very few that are easily observable with the naked eye.
Secondly, $\delta$\,Vel is known to have a moderate thermal infrared excess (\cite{aumann85}, \cite{su06}), and \emph{Spitzer} observations revealed a spectacular bow shock caused by the displacement of $\delta$\,Vel in a dense interstellar cloud (\cite{gaspar08}). Near and thermal infrared images of $\delta$\,Vel with the VLT/NACO (2.17\,$\mu$m) and VLT/VISIR ($8-13\,\mu$m) instruments did not show the presence of an excess of circumstellar origin, at a level of a few percents (\cite{kervella09}, Fig.~\ref{nacovisir}). This result indicates that these stars do not host a warm debris disk, and supports the conclusions of G\'asp\'ar et al.~(\cite{gaspar08}). It should be noted though that comparable observations in the thermal infrared domain using the Gemini South telescope showed a marginally resolved emission at $\lambda = 10.4\,\mu$m (\cite{moerchen10}).
Due to its very small angular separation (Sect.~\ref{eclipsing}), the eclipsing pair was first resolved using optical interferometry by Kellerer et al.~(\cite{kellerer07}) using the VLTI/VINCI instrument. Its composite spectrum is that of an early A star. Its distance was measured accurately by {\it Hipparcos} ($\pi = 40.50 \pm 0.4$\,mas; \cite{vanleeuwen07}).

\section{$\delta$\,Vel~Aa--Ab eclipsing binary properties and parallax of the system \label{eclipsing}}

We used two different approaches to determine the physical properties and orbital parameters of the eclipsing binary stars. Firstly, we combined only spectroscopy and photometry, and secondly, we added spatially resolved observations using interferometric data from the VLTI/AMBER instrument.

\subsection{Photometry and spectroscopy \label{pribulla_article}}

\begin{figure}[]
\centering
\includegraphics[height=5cm]{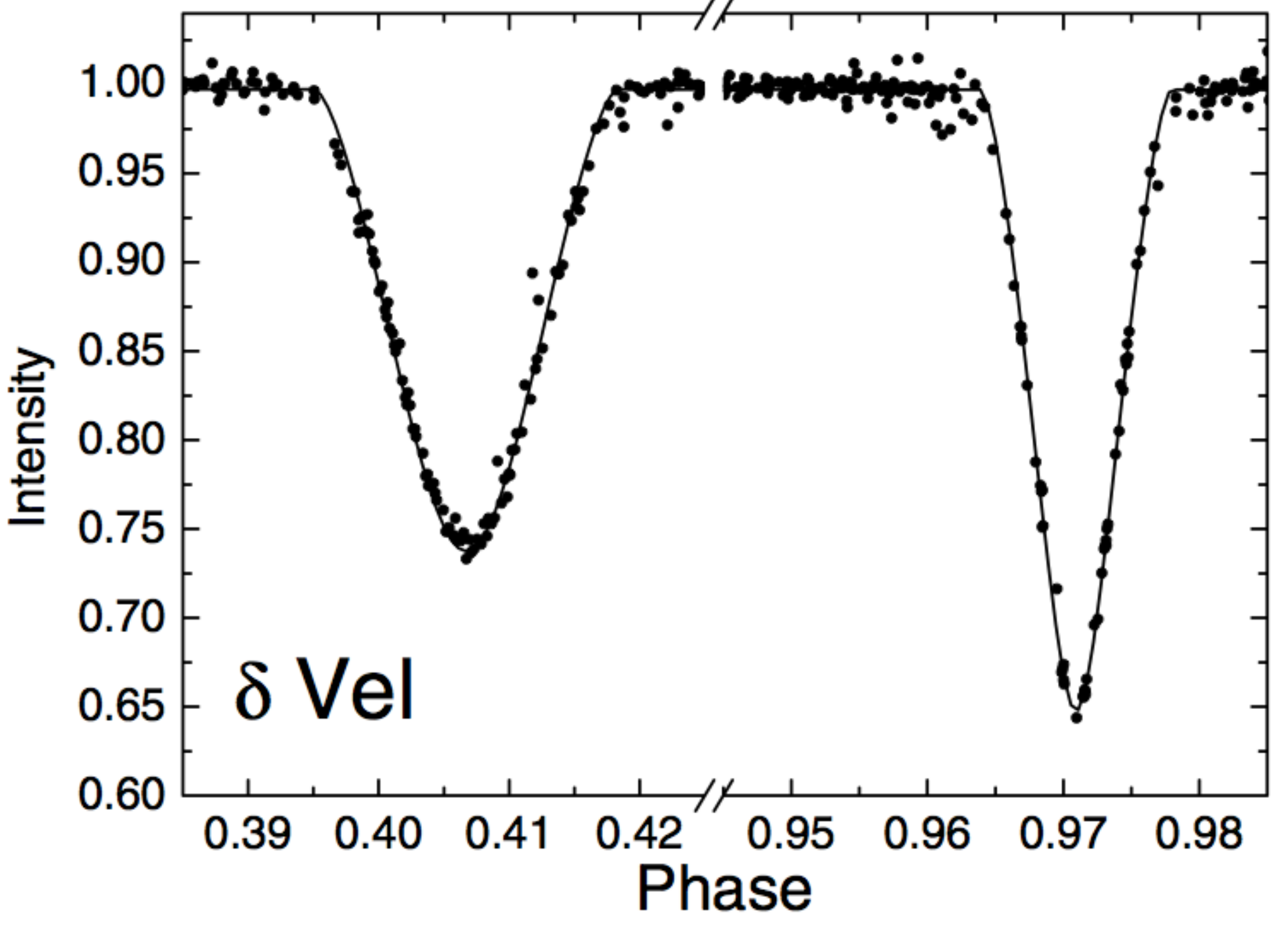} \hspace{8mm}
\includegraphics[height=4.5cm]{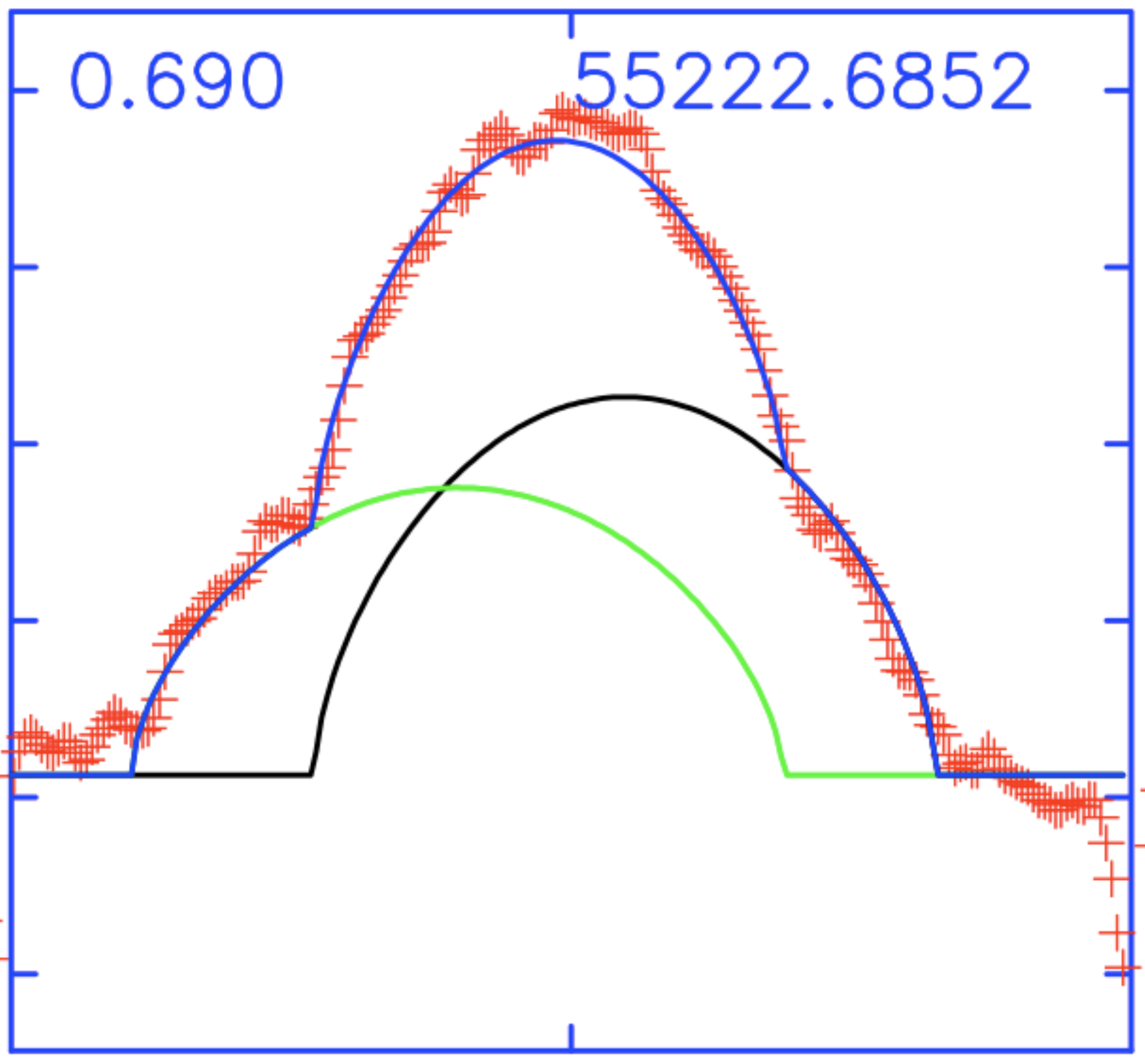}
\caption{{\it Left:} SMEI light curve of $\delta$\,Vel around the eclipses. {\it Right:} rotational profile fit to the observed broadening function at $\phi=0.690$ derived from the BESO spectroscopy. The different curves show the simulated profile of the Aa component (black), the simulated profile of Ab (green), the resulting total broadening function (blue) and the observed profile (red). These figures are taken from \cite{pribulla11}.  \label{lightcurve}}
\end{figure}

As described in \cite{pribulla11}, $\delta$\,Vel was observed using the SMEI instrument on board the Coriolis satellite. It produced more than 11\,000 integrated photometric measurements of the system over a period of 5.6\,years. Due to the long orbital period of the system, only about 350 measurements were obtained during the eclipses. The resulting detrended light curve, zoomed around the eclipses, is presented in Fig.~\ref{lightcurve} (left).
The spectroscopic observations were obtained using the BESO \'echelle spectrograph installed at the Cerro Armazones Observatory. Sixty-three spectra were collected between April 2009 and April 2010, covering a broad wavelength range (353--886\,nm). This blue part of the spectra was deconvolved using a high-resolution synthetic spectrum corresponding to $T_{\rm eff} = 9500$\,K, $\log g = 4.5$, solar metallicity , and a non-rotating star to obtain broadening functions (Fig.~\ref{lightcurve}, right). The extracted functions show clearly that $\delta$\,Vel Aab is indeed a SB2, but no signature of the B component was found in the spectra.
Combining the SMEI and BESO data, it was possible to derive the masses of the two main components of $\delta$\,Vel: $M(Aa) = 2.53 \pm 0.11\ M_\odot$, and $M(Ab) = 2.37 \pm 0.10\ M_\odot$, as well as their other physical parameters (see \cite{pribulla11} for details). It was in particular established that the two stars are fast rotators with $v \sin i$ values around 140\,km\,s$^{-1}$, and that the eclipses are not total.

\subsection{Interferometry \label{merand_article}}

\begin{figure}[ht]
\centering
\includegraphics[height=6cm]{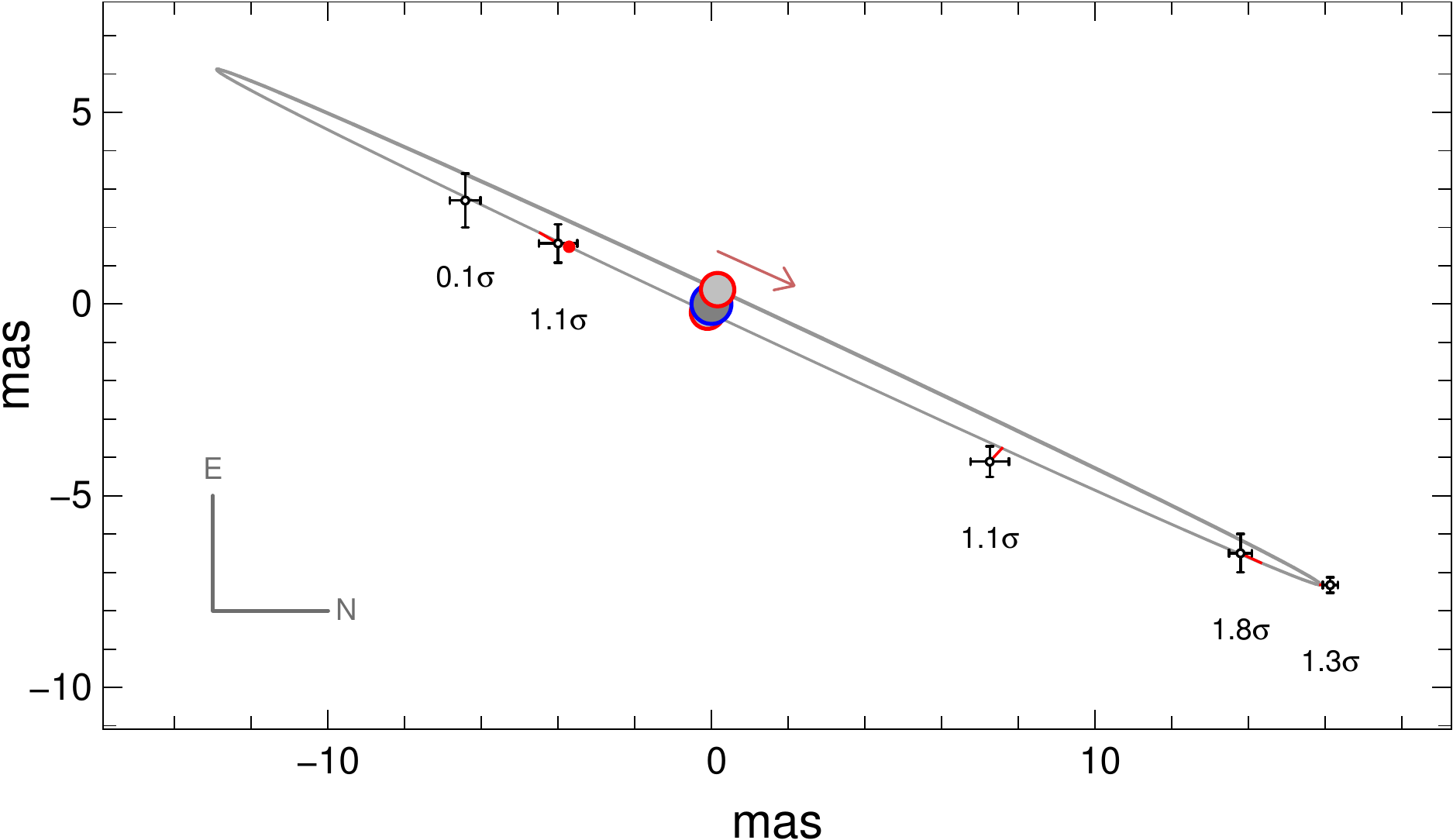} \hspace{3mm}
\caption{Visual orbit of the eclipsing binary $\delta$\,Vel~Aab as measured by interferometry with VLTI/AMBER (taken from \cite{merand11}) \label{Aab_orbit}}
\end{figure}

\begin{figure}[ht]
\centering
\includegraphics[height=6cm]{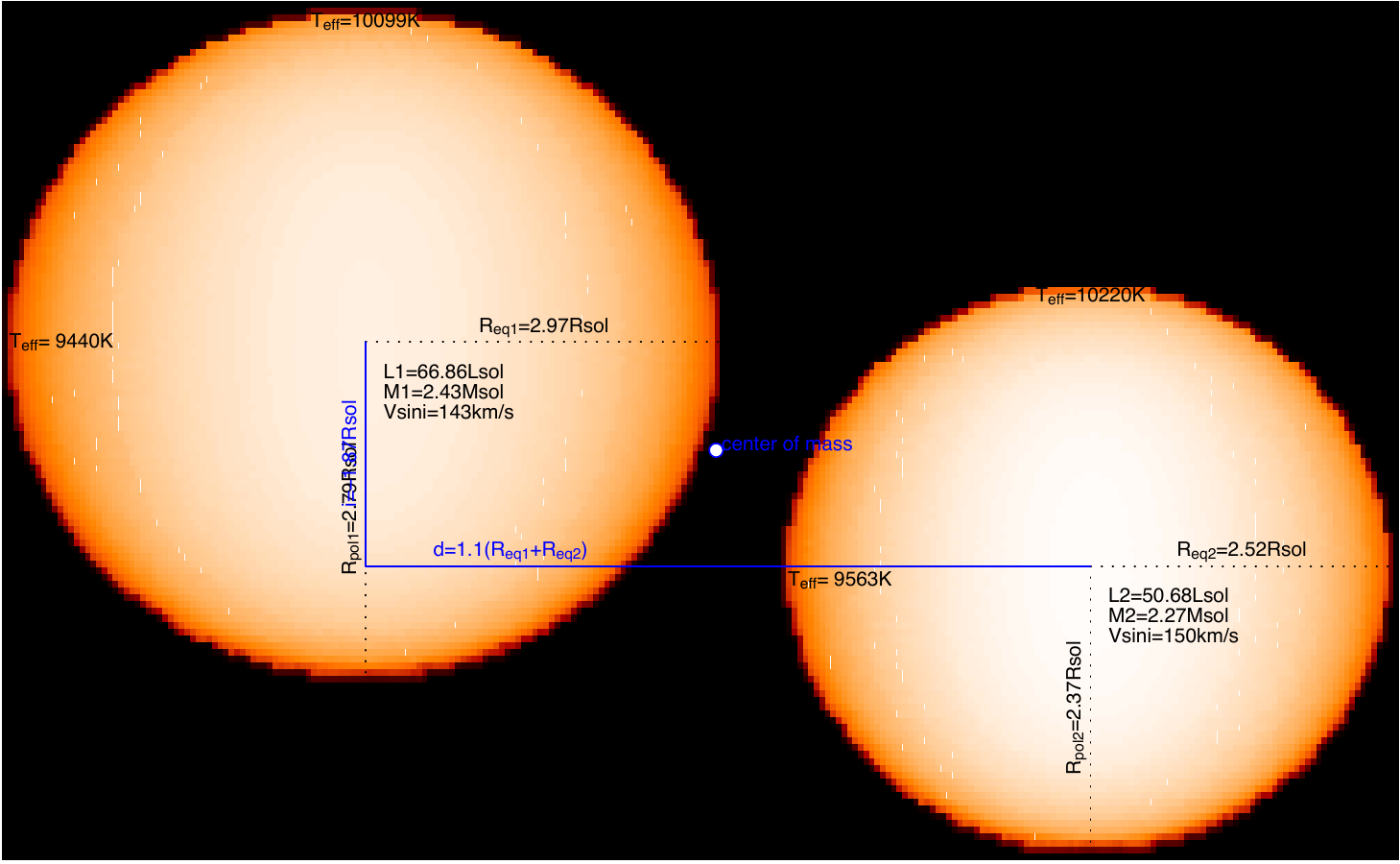}
\caption{Visual rendering of the Aab system close to the primary eclipse (taken from \cite{merand11}). \label{image1}}
\end{figure}

M\'erand et al.~(\cite{merand11}) collected interferometric observations of the $\delta$\,Vel Aab system using the AMBER instrument of the VLTI. This new observable allowed to retrieve the visual orbit of the eclipsing pair (Fig.~\ref{Aab_orbit}). Together with the existing photometry and spectroscopy, this gives the possibility to recover the distance of the system, independently of previous measurements. Moreover, this provides an improved precision for physical and orbital parameters of the two stars. The modeling of the photometric and radial velocity measurements of Aa-Ab pair was achieved using a self consistent method based on physical parameters (mass, radius, luminosity, rotational velocity). This model was used to reconstruct the aspect of the eclipsing binary (Fig.~\ref{image1}). It is interesting to remark in particular the effect of the fast rotation of the two components on the shape of the stars and the presence of equatorial darkening due to the Von Zeipel effect.
The two components appear to have similar masses of $2.43 \pm 0.02$ and $2.27 \pm 0.02\,M_\odot$. These values are in very good agreement with the results presented in Sect.~\ref{pribulla_article}. The parallax of the system is also derived, at $\pi = 39.8 \pm 0.4$\,mas, which is only $-1.2\,\sigma$ away from the Hipparcos value derived by van Leeuwen (\cite{vanleeuwen07}, $\pi_{\rm Hip} =40.5 \pm 0.4$\,mas).

\section{$\delta$\,Vel~AB visual pair orbital parameters}

\begin{figure}[ht]
\centering
\includegraphics[width=10cm]{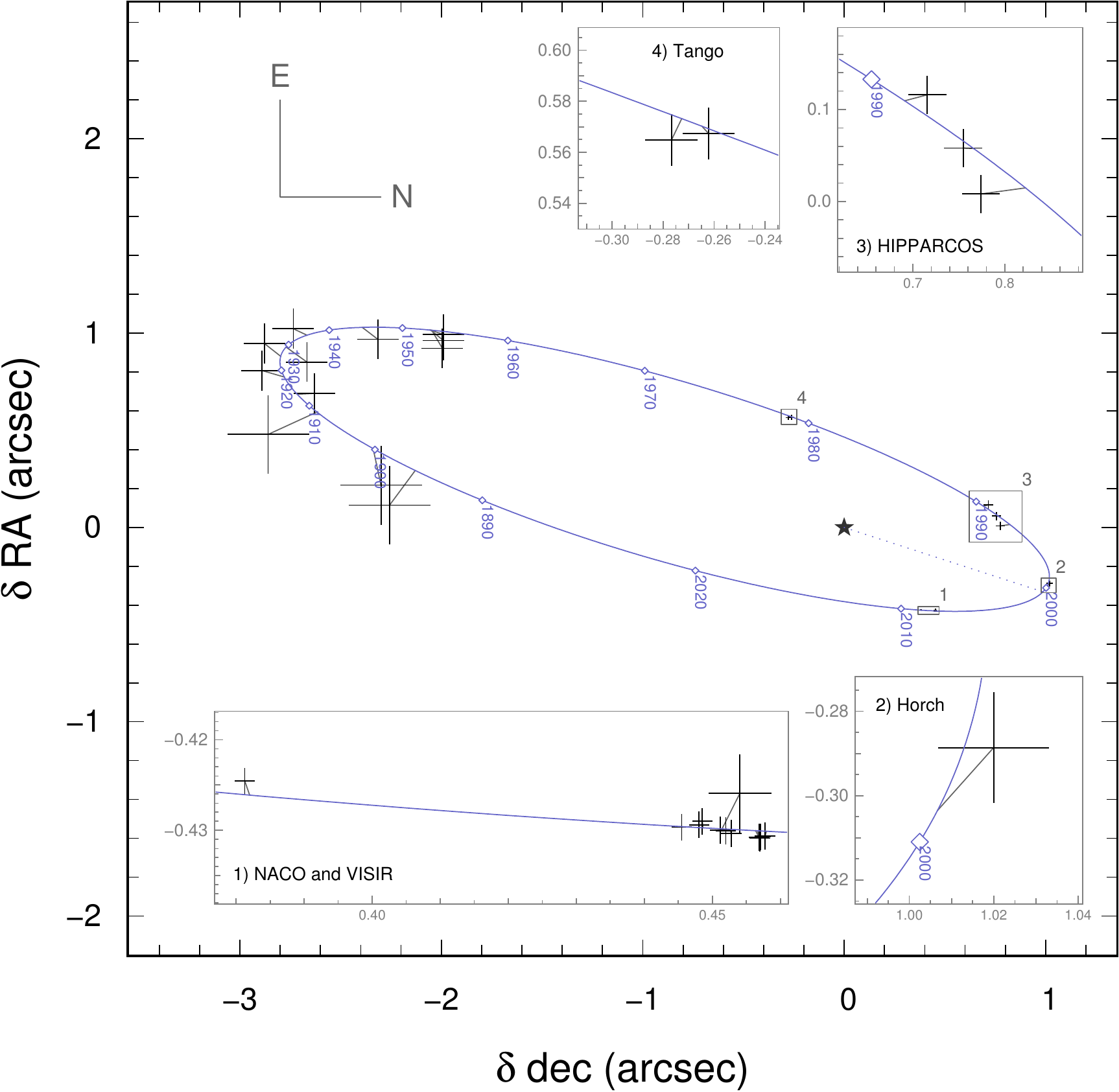}
\caption{Astrometry and adjusted orbit of $\delta$\,Vel~AB, from \cite{merand11} (the relevant references are given therein).  \label{ABorbit}}
\end{figure}

Over more than one century, the separation between $\delta$\,Vel~A and B has been decreasing at a rate which nicely matches the progression of the angular resolution of the successive generations of imaging instruments (visual observations, photography, electronic devices). This progression allowed a relatively regular monitoring of the visual orbit of the pair, down to the sub-arcsecond separations that occur around the periastron passage.  Thanks to the large aperture of the UT4 telescope of the VLT and the diffraction-limited angular resolution (\cite{merand11}), the NACO observations provide high-precision relative astrometry of the A-B pair. In addition to these recent astrometric measurements, we also took advantage of the historical astrometric positions assembled by Argyle et al.~(\cite{argyle02}), that include 17 epochs between 1895 and 1999. We adjusted the orbital parameters of the $\delta$\,Vel~A-B pair to the whole sample of astrometric data, and the result is presented graphically in Fig.~\ref{ABorbit}. The detailed orbital elements are listed in \cite{merand11}. Using the parallax determined from the eclipsing pair observations described in Sect.~\ref{eclipsing}, we obtain a total mass of $M(Aab + B) = 6.15 \pm 0.15_{\rm orbit} \pm 0.17_{\rm parallax}\ M_\odot$ for the three stars. The precision of this value is mainly limited by the accuracy of the parallax ($\pi = 39.8 \pm 0.4$\,mas).

\Conclusion

Thanks to its remarkable physical properties and proximity, $\delta$\,Vel emerged over the past decade as a particularly interesting triple stellar system. Its brightness makes it a perfect benchmark object for the application of a broad variety of observing techniques. Its very peculiar geometrical configuration is highly improbable, considering its orbital period. This also makes it an appealing target to search for extrasolar planets, that could be in orbit either around the individual eclipsing components Aa or Ab (although the stability of such planetary systems may be problematic) or around the Aab pair.

\vspace{3mm}

{\it Acknowledgements:}
This work received the support of PHASE, the high angular resolution partnership between
ONERA, Observatoire de Paris, CNRS and University Denis Diderot Paris 7. The research leading to these results has received funding
from the European Community's Seventh Framework Programme under Grant Agreement 226604. This work has also been supported by VEGA project
2/0094/11.

\end{document}